# MICROTUBULES: DYNAMICS, SOLITON WAVES, SOME ROLES IN THE CELL

Slobodan Zdravković

*Institut za nuklearne nauke Vinča, Univerzitet u Beogradu, Laboratorija za atomsku fiziku (040),11001 Beograd, Serbia. (szdjidji@vinca.rs)*

**ABSTRACT**
In the present paper we deal with nonlinear dynamics of microtubules (MTs). The structure and role of MTs in cells are explained. One model explaining MT dynamics is explained. Solutions of the crucial nonlinear differential equation depend on used mathematical procedures. Two of them, continuum and semi-discrete approximations, are explained. Finally, these solutions are shown and discussed. They are solitonic waves. Three different kinds of them are known in the moment. They are kink solitons, breathers and bell-type solitons.

**INTRODUCTION**
A cell is defined as eukaryotic if it has a membrane-bound nucleus**.** Such cells are generally larger and much more sophisticated than prokaryotic cells due to the many different types of specialized organelles present in most eukaryotic cells. Plant and animal cells are eukaryotic while bacteria cells are prokaryotic.

All eukaryotic cells produce two kinds of tubulin proteins. Alpha and beta tubulins spontaneously bind one another to form a functional subunit that we call a heterodimer, or a dimer for short. When intracellular conditions favour assembly, the dimers assemble into long structures called protofilaments (PFs). Microtubules (MTs) are usually formed of 13 PFs, as shown in Fig. 1. Hence, MTs are long cylindrical polymers whose lengths vary from a few hundred nanometers up to meters in long nerve axons [1]. Each dimer is an electric dipole whose mass is $m = 1.8 \times 10^{-22} \text{kg}$. Its length is $l = 8\text{nm}$, while the remaining two dimensions are $6.5\text{nm}$ and $4.6\text{nm}$ [2]. The component of its electric dipole moment in the direction of PF and charge displacement are: $p = 337\text{Debye} = 1.13 \times 10^{-27} \text{Cm}$ and $d \approx 4\text{nm}$, respectively [3].

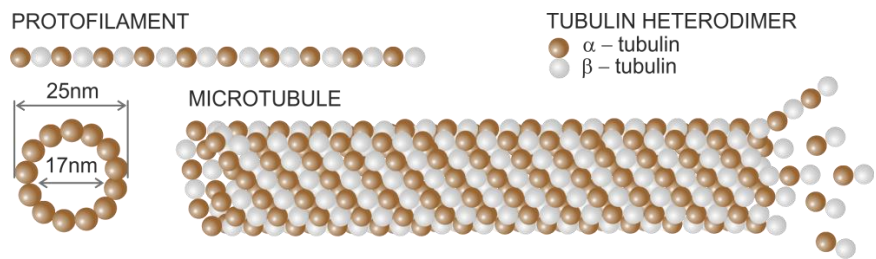

**Figure 1.** A tubulin dimers, a protofilament and a microtubule [4].

MTs are the major part of cytoskeleton (Fig. 2). They are long structures that spread between a nucleus and a cell membrane. MTs are involved in nucleic and cell division and organization of intracellular structure. They also serve as a network for motor proteins. This is shown in Fig. 3. There are two families of them called dyneins and kinesins. One can see, in Fig. 3, that the motor proteins carry different cargos, dynein towards the nucleus and kinesin towards the cell membrane. Motor proteins move with a velocity of $0.1-2\,\mu m/s$ [5]. For this activity they use the energy derived from repeated cycles of adenosine triphosphate (ATP) hydrolysis. One ATP molecule is hydrolyzed for each step of motor protein. The step of motor protein is $8\,nm$ distance, as this is nothing but one dimer. Energy released during ATP hydrolysis is about $14\,kcal/mol$, which corresponds to activation energy of the motor proteins [5].

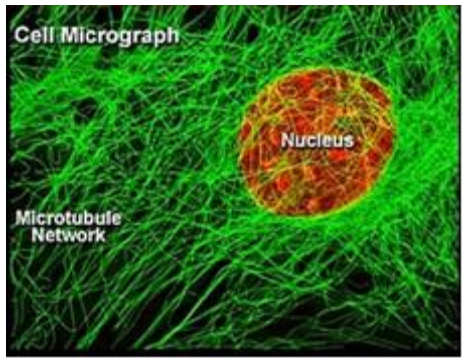 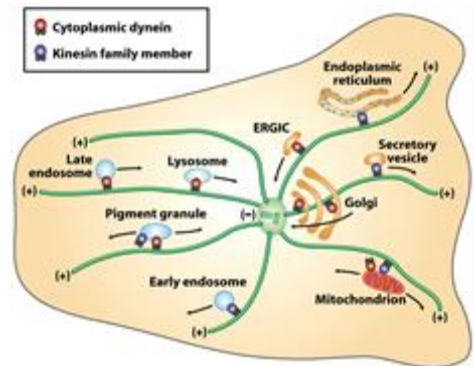

**Figure 2.** A nucleus and MT network in a eukaryotic cell [4].

**Figure 3.** Microtubule as a network for motor proteins in a eukaryotic cell [4].

MTs in non-neuronal cells are unstable structures. They undergo repeated cycles of cell division for which MTs disassemble and reassemble [1]. Populations of such MTs usually consist of some that are shrinking and some that are growing. Hence, they exhibit dynamic instability behaviour existing in phases of elongation or rapid shortening. MTs grow steadily and then shrink rapidly by loss of tubulin dimers at the plus end. The rapid disassembly is referred to as catastrophe. MTs elongate and shorten at velocities of $7.2\,\mu m/min$ and $17.3\,\mu m/min$, respectively [6]. In the interphase the half lifetime of individual unstable MT is $5-10\,min$ [5].

MTs existing in neuronal cells are uniquely stable and, consequently, neurons, once formed, don't divide [1]. This stability is crucial as there are evidences that neuronal MTs are responsible for processing, storage and transduction of biological information in a brain [1,7].

It was pointed out that MT was a hollow cylinder. This should not yield to a possible wrong conclusion that motor proteins move through it. Quite opposite, they "walk" along PFs carrying their cargos, as shown in Fig. 4.

Finally, meanings of the plus and the minus ends should be explained. MT polymerizes more quickly from the plus end, which is terminated by the $\beta$-subunit, i.e. $\beta$-monomer. The other end, growing more slowly, is known as the minus end, and is terminated by the $\alpha$-subunit. These dynamics ends do not correspond to electric ones. Namely, the dimers are electric dipoles and, consequently, the whole MT can be seen as a giant dipole, with its electric plus and minus ends. The electric plus end corresponds to the minus dynamics one and vice versa.

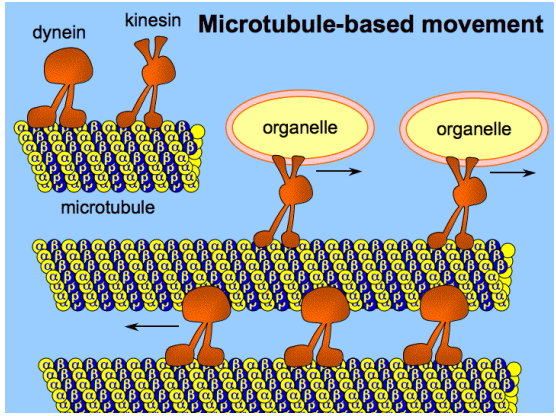 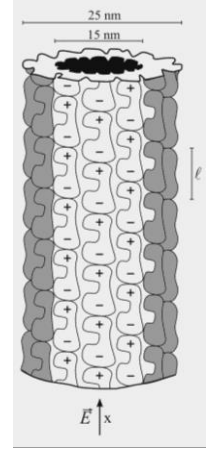

**Figure 4.** Motor proteins move along MTs carrying cargo [4].

**Figure 5.** Microtubule.

### NONLINEAR DYNAMICS OF MTs

To study dynamics of MTs we need an appropriate model. The dimers are electric dipoles and the whole MT can be regarded as ferroelectric. That was assumed for the first nonlinear model of MT [8].

The tubulin polymerization process involves two types of contacts between the dimers. These are head-to-tail binding of dimers, resulting in protofilaments, and interactions between parallel protofilaments, which complete the MT wall. Since the longitudinal contacts along PFs appear to be much stronger than those between adjacent PFs, we construct a simplified Hamiltonian of MT, which is, practically, Hamiltonian for a single PF only. However, the influence of the neighbouring PFs is taken into consideration through the electric field. Namely, each dimer exists in the electric field coming from the dimers belonging to all PFs. Also, the nearest neighbour approximation, very common in physics, is assumed.

There is one additional approximation. The dimers oscillate performing both radial oscillations around more than one axis and longitudinal ones along more than one direction. Hence, there are a few degrees of freedom but all existing models assume only one degree of freedom per dimer, though some attempts to use two degrees of freedom are in progress. Thus, according to the chosen coordinates describing dimers` oscillations we can talk about either radial or longitudinal model. The latter one we call as $u$-model [9]. Notice that the $u$–model assumes radial oscillations of the dimers but the coordinate $u$ is the projection of the top of the dimer on the direction of PF. There is a real longitudinal model assuming longitudinal displacements of the dimers that we call as $z$-model [10]. Both $u$- and $z$-models bring about equal differential equations but meanings of the coordinates $u$ and $z$ are different.

The second kind of the model is a radial one [11] and we call it as $\varphi$–model for short. Of course, the angle $\varphi$ determines the radial displacement of the dimer representing the angle between the dimer and the direction of PF.

The equation describing the $u$- and $z$-models comprises a term which does not exist in the one coming from the $\varphi$-model. Hence, it is more complicated and interesting and, in what follows, we concentrate on the *u*-model only.

Each model requires a simplified picture according to which we write Hamiltonian describing the physical system. Also, an elementary mass, i.e. the smallest particle whose internal structure is neglected, should be indicated. The *u*-model, as well as all the models mentioned above, assumes that the dimer is the elementary particle. The simplified picture, allowing us to write the Hamiltonian, could be the one in Fig. 5. The figure shows a segment of MT with three PFs clearly indicated.

The Hamiltonian for the *u*-model is

$$H = \sum_n \left[ \frac{m}{2}\dot{u}_n^2 + \frac{k}{2}(u_{n+1} - u_n)^2 - \frac{1}{2}Au_n^2 + \frac{1}{4}Bu_n^4 - Cu_n \right], \qquad C = QE, \quad (1)$$

where dot means the first derivative with respect to time, $m$ is a mass of the dimer, $k$ is an intra-dimer stiffness parameter, $Q > 0$ represents the excess charge within the dipole, $E > 0$ is internal electric field and the integer $n$ determines the position of the considered dimer in PF [8,9]. The first term represents the kinetic energy of the dimer. The second one is the potential energy of the chemical interaction between the neighbouring dimers belonging to the same PF, where, obviously, the nearest neighbour approximation is used, as explained above. The next two terms are called W-potential and they were introduced due to the fact that MT is ferroelectric and $A$ and $B$ are parameters that should be determined or, at least, estimated [8,9]. The very last term is coming from the fact that the dimer is the electric dipole existing in the field of all other dimers. The last three terms together can be called as the combined potential which looks like unsymmetrical W-potential.

Introducing generalized coordinates $q_n$ and $p_n$, defined as $q_n = u_n$ and $p_n = m\dot{u}_n$, and using well-known Hamilton's equations of motion we obtain the following discrete differential equation that should be solved

$$m\ddot{u}_n - k(u_{n+1} + u_{n-1} - 2u_n) - Au_n + Bu_n^3 - C = 0. \quad (2)$$

Therefore, nonlinear dynamics of MTs has been described by Eq. (2), which will be solved in the next section. Obviously, nonlinearity is coming from the 4th degree term in W-potential.

**SOLUTIONS OF EQ. (2)**

We are not going through all the tedious derivations. Instead, two known approaches will be explained briefly. It is interesting that the final results depend on applied mathematical procedures.

Some approximations used in derivation of Eq. (2) were introduced above. We now explain the two mathematical methods for solving this equation. Practically, these two approaches are two approximations. In other words, the approximations mentioned above were important for derivation of Eq. (2), while the mathematical ones are important for its solutions.

The first mathematical approximation that we explain is a continuum approximation $u_n(t) \to u(x,t)$, which allows a series expansion of the terms

$u_{n\pm1}$. This brings about the following continuum dynamical equation of motion

$$m\frac{\partial^2 u}{\partial t^2} - kl^2\frac{\partial^2 u}{\partial x^2} - qE - Au + Bu^3 + \gamma\frac{\partial u}{\partial t} = 0.  \qquad (3)$$

The last term is coming from the introduced viscosity force $F_v = -\gamma\dot{u}$, where $\gamma$ is a viscosity coefficient [8].

The remaining mathematical approximation is a semi-discrete one [9,12,13]. We assume small oscillations $u_n(t) = \varepsilon\,\Phi_n(t)$, where $\varepsilon << 1$, and look for wave solution which is a modulated wave. Its envelope is continual, while its carrier component is discrete. Such solution can be written as

$$\Phi_n(t) = F(\xi)e^{i\theta_n} + \varepsilon\,F_0(\xi) + \varepsilon\,F_2(\xi)e^{i2\theta_n} + \text{cc} + \text{O}(\varepsilon^2), \qquad (4)$$

where $\xi = (\varepsilon nl, \varepsilon t)$, $\theta_n = nql - \omega t$, $\omega$ is the optical frequency of the linear approximation, $q = 2\pi/\lambda > 0$ is the wave number, cc represents complex conjugate terms and the function $F_0$ is real. The dimers length $l$, mentioned above, is nothing but a period of one dimensional crystal lattice. Obviously, the continuous function $F$ represents an envelope, while $e^{i\theta_n}$, including discreteness, is the carrier component. The first term in the expansion (4) is a leading one, while the remaining ones, multiplied by $\varepsilon$, are higher harmonics, representing a certain correction. Notice that the parameter $\varepsilon$ exists in the function $F$ but does not in $e^{i\theta_n}$. This is so because the frequency of the carrier wave is much higher than the frequency of the envelope and we need two time scales, $t$ and $\varepsilon t$, for those two functions. Of course, the same holds for the coordinate scales.

This method is very tedious and we do not go through all the derivations. The complete procedure and important explanations can be found in Ref. [14]. The point is that the functions $F_0(\xi)$ and $F_2(\xi)$, existing in Eq. (4), can be expressed through $F(\xi)$. In this particular case they turned out to be constants [9]. Hence, we only need the equation for the function $F(\xi)$ and this is a well-known solvable nonlinear Schrödinger equation (NLSE)

$$iF_\tau + PF_{SS} + Q\left|F\right|^2 F = 0, \qquad (5)$$

where the dispersion coefficient $P$ and the coefficient of nonlinearity $Q$ are explained in the aforementioned references, as well as the coordinates $\tau$ and $S$.

Therefore, the crucial Eq. (2), describing nonlinear dynamics of MTs, has been transformed into Eqs. (3) and (5). The final step is the solutions of these two equations. This is a topic of the next section.

**SOLITONIC WAVES IN MICROTUBULES**
The very first observation of a soliton was made in 1834 by the hydrodynamic engineer John Scott Russel [13]. He was riding his horse, while a pair of horses drew a boat along a narrow channel. When the boat was suddenly stopped he noticed an interesting wave moving along the channel. That was a smooth and well defined heap of water, which moved along the channel without change of form or diminution of speed [13]. The wave was so stable that the engineer followed it about one or two miles. He called this phenomenon the wave of translation.

Understanding of this interesting observation is nothing but a derivation of an equation whose solution is the observed wave. This equation is now called as KdV equation according to the initials of its authors Korteweg and de Vries, who derived it in 1895. More than 20 years earlier this equation existed in an implicit form in research of Boussinesq.

Nonlinear equations and solitonic waves are nowadays very common in many branches of sciences. Being a non-linear, it became fashionable. Now we return to nonlinear biophysics, i.e. to Eqs. (3) and (5). The former one is a partial differential equation (PDE). Generally, PDEs cannot be easily solved. Hopefully, Eq. (3) can be transformed into an ordinary differential equation (ODE) introducing a unified variable $\xi$ defined as $\xi \equiv \kappa x - \omega t$, where $\kappa$ and $\omega$ are constants. Substitution of $x$ and $t$ by $\xi$ transforms Eq. (3) into the following ODE

$$\alpha \psi'' - \rho \psi' - \psi + \psi^3 - \sigma = 0, \qquad (6)$$

where a dimensionless function $\psi$ has been introduced through the relation $u = \sqrt{A/B}\,\psi$. The values of the parameters $\alpha$, $\rho$ and $\sigma$ can be found in aforementioned references. It is important to keep in mind that $\rho$ and $\sigma$ are proportional to the viscosity coefficients and electric field, respectively, while $\alpha$ can be determined.

Eq. (6) has been solved using a couple of mathematical procedures such as standard procedure [8], modified extended tanh-function method, procedure based on Jacobian elliptic functions, method of factorization and the simplest equation method (SEM). Beside a couple of diverging functions all these procedures bring about the same solutions having physical sense. These are three kink solitons, shown in Fig. 6. Of course, the function $\psi(\xi)$ is the solution of Eq. (6). These solitons are kink-solitons or kinks ($\psi_3$) and antikink-solitons ($\psi_1$ and $\psi_2$). It is common to call both of them as kinks for short. If we look at any of them we see that the kinks represent transition between two asymptotic states ($\xi \to \pm\infty$). In other words, orientation of the dimers is changed. Regarding the present figure the transition occurs in the approximate interval of $\xi \in (-6,+6)$. This interval is moving along MT.

The most general method SEM brings about infinitely many parallel lines corresponding to all the three functions in Fig. 6. Of course, this is of mathematical interest but all those functions have the same physical meaning.

The advantage of SEM method over the remaining ones was demonstrated in Ref. [15]. A completely new solution was obtained. This is a bell-type soliton, shown in Fig. 7. This solution of Eq. (6) can be obtained only for $\rho = 0$, i.e. when viscosity is neglected.

Therefore, two completely different solutions of Eq. (6) have been derived. A final step is derivation of Eq. (5). Even though this is PDE its solution exists [13,14]. Hence, we know $F$, the functions $F_0$ and $F_2$, existing in Eq. (4), can be expressed through $F$ and all this brings about the final expression for $u$. This interesting solution is

$$u_n(t) = A_0 \text{sech}\left(\frac{nl - V_e t}{L}\right)\cos(\Theta nl - \Omega t) - \frac{C}{A}, \qquad (7)$$

where the expressions for all the parameters can be found in Ref. [9]. The hyperbolic function represents the envelope, while the remaining one is the carrier wave. This is shown in Fig. 8. Obviously, this is a localized modulated wave usually called as breather. We see that its width is about 200nm, which means that it covers about 25 dimers.

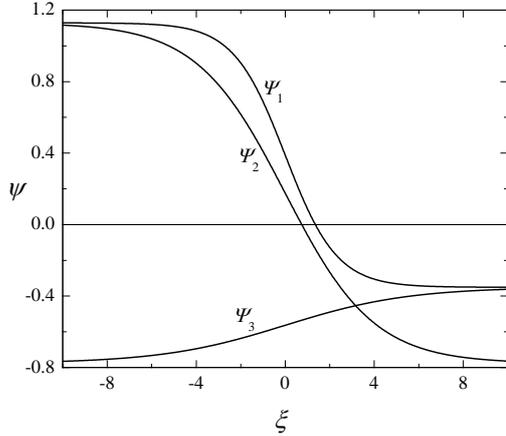

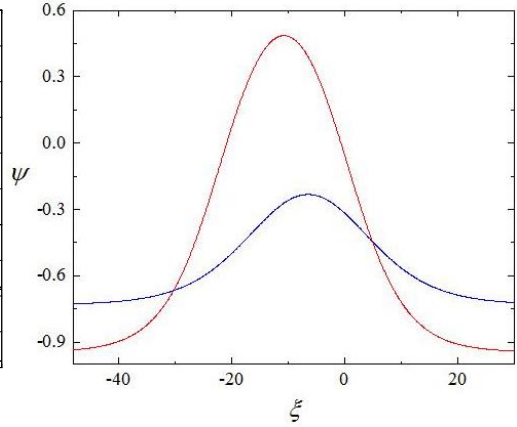

**Figure 6.** The functions $\psi(\xi)$ for $\rho = 2$ and $\sigma = 0.31$.

**Figure 7.** A bell-type soliton for $\sigma = 0.34$ (blue) and $\sigma = 0.1$ (red).

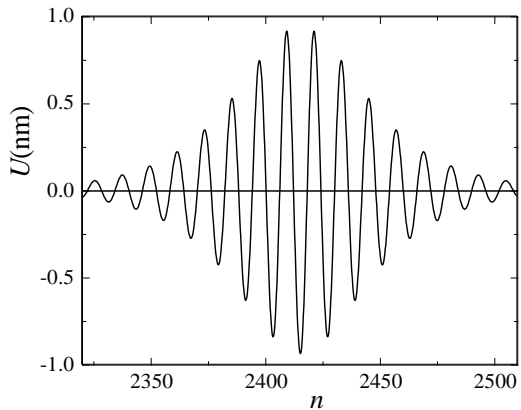

As a conclusion we can say that all the three known kinds of solitonic waves have been found in MTs. All these analytical results have been numerically supported.

**Figure 8.**
A localized modulated soliton (breather)

**CONCLUSION AND FUTURE RESEARCH**
The three different solitons have been predicted as possible candidates for information carriers along MTs. They are shown in Figs. 6-8. A crucial question is their stability, which is very important task that should be performed.

A weak point in the *u*-model is the last term in Eq. (1). The $\varphi$−model was an attempt to overcome the problem. A better potential energy $U(\varphi) = -\vec{p}\cdot\vec{E} = -qdE\cos\varphi$ was introduced instead but the W-potential has been lost. Hence, research in progress has been carried out with a goal to create a general model which will include both $U(\varphi)$ and W-potential.

Let us return to Eq. (1) and Fig. 5. According to the latter, one would conclude that all the dimers are positioned in the direction of PFs. Of course, this is a simplified picture. The last three terms in Hamiltonian given by Eq. (1) represent the nonsymmetrical W-potential. The graph of this function has two minima. This means that there are two possible orientations of the dimers, i.e. two possible angles between them and PFs. According to the dimensions of the dimers we can expect that it should be possible to measure these angles. Such experiments would prove or disapprove the theoretical expectation regarding W-potential. In case that the used potential has been a good choice the measured angles would improve theory a lot.

Therefore, stability analysis, the general model and the experimental verifications are new tasks that should be performed in near future.

*Acknowledgement*
This work was supported by funds from Serbian Ministry of Education, Sciences and Technological Development (grant No. III45010).